# Particle Physics Aspects of Antihydrogen Studies with ALPHA at CERN[*]

M.C. Fujiwara[a], G. B. Andresen[b], W. Bertsche[c], P.D. Bowe[b], C.C. Bray[d], E. Butler[c], C. L. Cesar[e], S. Chapman[d], M. Charlton[c], J. Fajans[d], R. Funakoshi[f], D.R. Gill[a], J.S. Hangst[b], W.N. Hardy[g], R.S. Hayano[f], M.E. Hayden[h], A.J. Humphries[c], R. Hydomako[i], M.J. Jenkins[c], L.V. Jørgensen[c], L. Kurchaninov[a], W. Lai[a], R. Lambo[e], N. Madsen[c], P. Nolan[j], K. Olchanski[a], A. Olin[a], A. Povilus[d], P. Pusa[j], F. Robicheaux[k], E. Sarid[l], S. Seif El Nasr[g], D.M. Silveira[e], J.W. Storey[a], R.I. Thompson[i], D.P. van der Werf[c], L. Wasilenko[a], J.S. Wurtele[d], and Y. Yamazaki[m]

(ALPHA Collaboration)

[a]*TRIUMF, 4004 Wesbrook Mall, Vancouver, BC, V6T 2A3, Canada*
[b]*Department of Physics and Astronomy, Aarhus University, DK-8000 Aarhus C, Denmark*
[c]*Department of Physics, Swansea University, Swansea SA2 8PP, United Kingdom*
[d]*Department of Physics, University of California at Berkeley, Berkeley, CA 94720-7300, USA*
[e]*Instituto de Fisica, Universidade Federal do Rio de Janeiro, Rio de Janeiro 21941-972, Brazil*
[f]*Department of Physics, University of Tokyo, Tokyo 113-0033, Japan*
[g]*Department of Physics and Astronomy, University of British Columbia, Vancouver BC, Canada*
[h]*Department of Physics, Simon Fraser University, Burnaby BC, V5A 1S6, Canada*
[i]*Department of Physics and Astronomy, University of Calgary, Calgary AB, T2N 1N4, Canada*
[j]*Department of Physics, University of Liverpool, Liverpool L69 7ZE, United Kingdom*
[l]*Department of Physics, Auburn University, Auburn, AL 36849-5311, USA*
*Atomic Physics Laboratory, RIKEN, Saitama 351-0198, Japan*

**Abstract.** We discuss aspects of antihydrogen studies, that relate to particle physics ideas and techniques, within the context of the ALPHA experiment at CERN's Antiproton Decelerator facility. We review the fundamental physics motivations for antihydrogen studies, and their potential physics reach. We argue that initial spectroscopy measurements, once antihydrogen is trapped, could provide competitive tests of CPT, possibly probing physics at the Planck Scale. We discuss some of the particle detection techniques used in ALPHA. Preliminary results from commissioning studies of a partial system of the ALPHA Si vertex detector are presented, the results of which highlight the power of annihilation vertex detection capability in antihydrogen studies.

---

[*] Invited talk at Pbar08 – Workshop on Cold Antimatter Plasmas and Application to Fundamental Physics.

# INTRODUCTION

One of the prime goals of antihydrogen research is to study symmetry between matter and antimatter. Given that atomic hydrogen is one of the best studied systems in physics, a comparison of hydrogen and antihydrogen properties belongs to the class of experiments that provide the foundations of modern physics. Cold antihydrogen atoms were first produced by the ATHENA [1] and ATRAP [2] experiments in 2002 at CERN's Antiproton Decelerator (AD), currently the world's only source of low energy antiprotons. These anti-atoms, while nearly at rest, were not trapped and hence annihilated on the apparatus walls shortly after production. The next major step in the field is stable trapping of antihydrogen atoms, and this is the short-term goal of ALPHA, the successor of ATHENA.

The ultimate goal of performing fundamental tests is ambitious: numerous technical developments are required in order to produce, detect, trap, cool and interrogate anti-atoms. This requires us to drive progress in atomic, plasma, and ion trap physics in regimes previously unexplored, even with matter particles. Since its start in 2006, ALPHA has been making aggressive progress, already producing several publications [3,4,5,6,7]. See Ref. [8] for a review of the early results.

In this article, we will discuss particle physics aspects of antihydrogen studies, particularly focusing on the physics motivations and particle detection techniques.

# FUNDAMENTAL PHYSICS MOTIVATIONS

In this section, we discuss the scientific case for antihydrogen research in general and ALPHA in particular, from the viewpoint of fundamental subatomic physics. Plasma and trap physics related interests in antihydrogen studies are not described here, but see, e.g., Ref. [9] for discussions of related phenomena.

## CPT symmetry

Our belief in CPT invariance is largely theoretical, based on the remarkable success of quantum field theory. Whether CPT is in fact an exact symmetry is a question that should be tested by experiment. Recall that Nature has given us a list of symmetries that are fundamental, yet broken: parity, time-reversal, electroweak, chiral, and perhaps supersymmetry. Precision comparisons of antihydrogen with hydrogen atoms could provide some of the most stringent direct tests of CPT, with the possibility of probing the energy scale beyond the Planck scale.

The CPT theorem [10] guarantees the invariance of CPT in quantum field theories in a flat space time, with mild assumptions including Lorentz invariance, locality, unitary and the spin statistics connection. These assumptions may not be valid in quantum gravity, string theory, or theories with large extra dimensions.

CPT violation could also have implications in cosmology. Baryon asymmetry in the Universe is usually associated with the famous Sakharov condition: (1) Baryon number violation, (2) C and CP violation, and (3) that these occur out of thermal equilibrium. It is, however, possible to generate baryon number excess without (2) and (3), if CPT is violated [11,12]. In fact, an $O(10^{-6})$ difference in top and anti-top quark mass can generate the observed baryon asymmetry [13]. Note that the (anti)top quark mass is currently only known to the 1% level. Another important, yet challenging, goal is a test of the gravitational interaction between matter and antimatter, for which there exist no direct experimental data. Such measurements will test the Weak Equivalence Principle of general relativity.

**TABLE 1**. Direct tests of CPT via particle-antiparticle comparisons [14].

| Particle, CPT quantity | Relative precision | $\Delta$m in energy |
|---|---|---|
| e- e+ mass | $0.8 \times 10^{-8}$ | $4 \times 10^{-12}$ GeV |
| $K^0 \overline{K^0}$ mass※ | $10^{-18}$※ | $5 \times 10^{-19}$ GeV※ |
| $p \overline{p}$ mass | $2 \times 10^{-9}$ | $2 \times 10^{-9}$ GeV |
| e- e+ g-2 | $2 \times 10^{-9}$ | |
| $\mu$- $\mu$+ g-2 | $0.7 \times 10^{-6}$ | |
| $p \overline{p}$ q/m | $0.9 \times 10^{-10}$ | |
| $p \overline{p}$ magnetic moment | $3 \times 10^{-3}$ | |

※See text for discussion of the kaon test.

A theory recently proposed by Kostelecky and co-workers, the so-called Standard Model Extension (SME) [15] has generated much attention with regard to CPT and Lorentz violation tests. According to the SME, a large parameter space for CPT violation can be explored by searching for Lorentz violation signals with normal matter systems, and impressive progress is being made in this area [16]. Direct comparisons of matter-antimatter systems, on the other hand, give model-independent tests of CPT invariance. Table 1 shows some of the existing direct tests of CPT, with emphasis on the mass measurements.

It should be noted that meson-antimeson systems ($q\bar{q} - \bar{q}q$) are inherently different from baryon-antibaryon ($qqq - \bar{q}\bar{q}\bar{q}$) or lepton-antilepton ($l - \bar{l}$) systems, and could have different patterns of symmetry violation. Given the fundamental importance of CPT symmetry, it should be tested in all particle sectors. Furthermore the interpretation of the $K^0 - \overline{K^0}$ results as a test of CPT is controversial. The impressive quoted precision is derived with some theoretical assumptions, and Kobayashi and Sanda [17] have put forward an argument that the CPT violating interaction is tested only at much reduced level (currently ~$10^{-5}$ [18]). Bigi has also questioned the quoted relative precision of $10^{-18}$ for similar reasons [19].

In addition to the relative precision, one can consider the absolute energy precision as an alternative figure of merit. In precision measurements, typical energy corrections from short distance effects may have the form [20]:

$$\Delta Energy \sim \frac{m^{n+1}}{\Lambda_{NP}^{n}} \qquad (1)$$

where *n* depends on the theory and is typically > 0, and *m* is the characteristic energy scale of the system, while $\Lambda_{NP}$ is the high energy scale applicable to the CPT violating new physics.[†] To give a scale, for $n = 1$, $m \sim$ GeV, and $\Lambda_{NP} \sim m_{Pl}$ (the Planck scale $\sim 10^{19}$ GeV), we may expect the CPT violating energy shift at $\Delta Energy \sim 10^{-19}$ GeV level. The $K^0 - \overline{K^0}$ comparison, taken at face value, is constrained only to $\sim 100$ kHz in frequency or $5 \times 10^{-19}$ GeV in terms of the absolute energy scale. In this respect, antihydrogen experiments could compete favorably with the kaon and other direct tests of CPT, the details of which now follow.

## Physics Reach with Antihydrogen Laser Spectroscopy

As previously mentioned, atomic hydrogen is one of the most precisely studied simple systems in physics. The two photon transition between the 1s and 2s states is an ideal system for precision spectroscopy, due to its narrow intrinsic line width ($\Delta\nu/\nu = 4 \times 10^{-16}$), as well as the possibility for cancellation of the 1st order Doppler broadening when counter propagating photons are employed (the well-established "Doppler-free spectroscopy" techniques). Further, spectral lines collected at sufficiently high signal-to-noise level can be fit to known spectral shapes, allowing determination of transition line frequencies to well below natural linewidths. In principle, if this approach could be used to "split" the line by a factor of 500, precisions approaching $10^{-18}$ would be achievable. Obviously, there are many technical challenges before this ultimate precision can be realized. Nonetheless, progress in the measurement of this transition in atomic hydrogen has been remarkable, and the current relative precision is at a level of $2 \times 10^{-14}$ with an uncertainty of $\sim 50$ Hz [21].

While the ultimate precision of $10^{-18}$ may become possible one day, what precision can we achieve with antihydrogen in the short term? Clearly, because of the scarcity of the anti-atoms, the technological bottle-neck is in achieving stable trapping. In 1993 Hänsch and Zimmermann estimated that 1000 trapped antihydrogen atoms would be required to achieve $10^{-12}$ precision in 1s-2s spectroscopy [22]. In fact, this level of line width has been observed with trapped hydrogen atoms by a member of the ALPHA collaboration in 1996 [23]. With the progress in producing continuous Lyman alpha lasers, Walz et al. [24] argued that even fewer atoms would be needed to make precision spectroscopy by using the shelving scheme, i.e., repeating many times the transitions of the same atom. The precision at this level would be limited in part by the Zeeman effect in the trapping magnetic field. If laser cooling, which has been demonstrated for trapped hydrogen atoms [25], can be applied to antihydrogen, it would dramatically reduce the influence of the Zeeman effect, hence improving the precision. Introduction of a high power laser into an antihydrogen production apparatus has already been achieved [26], hence the technical bottle-neck for antihydrogen laser cooling is the stable trapping of antihydrogen. Many other developments relevant for antihydrogen spectroscopy are being reported in the literature. (For example, 2 photon laser cooling [27], coil-gun slowing [28], single atom cooling [29], Rydberg trapping [30], to name just a few).

---

[†] Since direct CPT tests challenge the entire framework of (effective) field theory, one has to be cautious in applying the language of field theory as in Eq. 1.

An initial $10^{-12}$ measurement of the 1s-2s transition would be already competitive with other CPT tests involving leptons and baryons (Table 1). In particular, it would constitute a 4 order of magnitude improvement in the equality of the electron and the positron mass[‡]. In terms of absolute precision, $\Delta\nu$ ~ few kHz corresponds to an energy sensitivity of $\Delta E$ ~ $10^{-20}$ GeV.

Table 2 illustrates the physics reach of the initial spectroscopy measurement with trapped antihydrogen. These initial measurements could already reach the Planck suppressed $1/M_{pl}$ sensitivity. Further improvements are envisioned with existing and emerging atomic techniques.

**TABLE 2.** Expected precision of initial 1s- 2s measurement and the physics reach

| Relative precision $\Delta\nu/\nu$ | Frequency precision $\Delta\nu$ | Absolute energy sensitivity $\Delta E$ | CPT quantity | Possible improvement factor |
|---|---|---|---|---|
| $10^{-12}$ | few $10^3$ Hz | $10^{-20}$ GeV | e- e+ mass ratio[ii] | $10^4$ |

## Antihydrogen Hyperfine Splitting

We are developing a technique for preliminary microwave spectroscopy of antihydrogen hyperfine splitting in a high magnetic field, based on the present ALPHA apparatus [8]. Hyperfine measurement in a low field has been considered by the ASACUSA experiment [31]. The hyperfine fine splitting of atomic hydrogen, the famous 21 cm line, is very well measured to the $7 \times 10^{-13}$ level with an absolute precision of 1 mHz. The magnetic moment of antiprotons is currently only known to 0.3%. The deviation of the (anti)proton magnetic moment from the Dirac value reflects its internal structure, and hence its value is possibly more sensitive to short-distance effects than 1s-2s transitions. In addition, *within the SME model*, the sensitivity to the CPT violating effects are enhanced by $O(1/\alpha^2)$ ~ $10^4$, compared to the 1s-2s case. Hyperfine splitting can therefore provide a powerful complementary test of CPT.

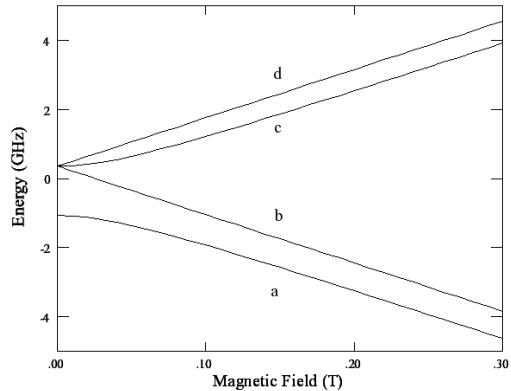

**FIGURE 1.** Breit-Rabi diagram for energy levels of (anti)hydrogen atoms in a magnetic field.

---

[‡] The charge equality $|q_e| = |q_{e+}| = |q_p| = |q_{\bar{p}}|$ is assumed to derive the e- e+ mass ratio from the hydrogen-antihydrogen comparison of the 1s-2s level. Alternatively, limits on the charges can be obtained.

The energy levels of (anti)hydrogen atoms in a magnetic field, given by the Breit-Rabi formula, are shown in Fig. 1. Microwave transitions from low field seeking, trappable states (c, d in Fig. 1) to high field seeking un-trappable states (a, b) would lead to ejection of the antihydrogen atom from the trap. The transition can be detected, with nearly 100% efficiency, via annihilation signals on the trap wall measured by ALPHA's position sensitive vertex detector (see below). An onset of this signal is expected to occur when the frequency of an applied microwave is varied such that the anti-atoms come into resonance. Each transition frequency is sensitive to a different combination of fundamental parameters. For example, in the high-field limit, the transitions between the states (d→ a) and (c→b) are given by [8]:

$$\nu_{ad} = \frac{\Delta \nu_H}{2} + \frac{\mu_B g_{e+}}{h} B, \text{ and}$$
$$\nu_{bc} = -\frac{\Delta \nu_H}{2} + \frac{\mu_B g_{e+}}{h} B. \quad (2)$$

Here $\Delta\nu_H$ is the antihydrogen hyperfine splitting at zero field, $\mu_B$ the Bohr magneton, and $g_{e+}$ the positron g-factor in the antihydrogen atom. The difference $(\nu_{ad} - \nu_{bc}) = \Delta\nu_H$ will give a measure of the antihydrogen hyperfine splitting, which is proportional in leading order to the antiproton magnetic moment. The sum $(\nu_{ad} + \nu_{bc})$ will give the value of the positron bound state g-factor. The resonant microwave transitions as in Eq. (2) may be one of the first spectroscopic signatures of trapped antihydrogen. Given that the antiproton magnetic moment is currently known only to 0.3%, a measurement of $\Delta\nu_H$ with an initial precision of 0.1% would already constitute a significant CPT test of the magnetic properties of an anti-baryon. A $10^{-4}$ measurement would reach an absolute energy sensitivity of ~$10^{-18}$ GeV, competitive with the neutral kaon test. Dramatic improvements beyond these proof-of-principle measurements can be envisioned, when combined with existing and emerging cooling and trapping techniques.

## Antimatter Gravity

Finally, the gravitational influence on antimatter could be observable with antihydrogen atoms laser cooled to mK.[§] For a vertical trap, the antihydrogen density would have the form ~ $\exp(-M \bar{g} h / kT)$, where $M$ is the antihydrogen gravitational mass, $\bar{g}$ the gravitational acceleration for antimatter and $T$ the antihydrogen temperature [32]. For $T = 1$ mK, we have a typical height of $h \sim 85$ cm. Hence with a few mK antihydrogen, we would know at least the sign $\bar{g}$ of in a reasonably sized apparatus. Other measurements of antimatter gravity have been proposed at the AD [33].

---

[§] The Doppler limit is 2.4 mK, and the recoil limit 1.3 mK, for Lyman alpha cooling.

# PARTICLE DETECTION IN ALPHA

In this section, we discuss some of the particle detection systems used in the ALPHA experiment. There are two distinct kinds of particle detection that are relevant. One is *counting* of antiproton annihilation events, and the other is *imaging* of antiproton annihilations. A unique feature of antiparticle studies (as opposed to normal particles) is the high efficiency for its detection via annihilations, which allows sensitive studies of important processes such as antihydrogen production and antiproton losses due to non-uniform magnetic fields.

## External Annihilation Detector

The annihilation counting technique is often used to count the number of trapped antiprotons. When trapped antiprotons are released to collide with the apparatus walls, an annihilation event produces several charged particles (mostly pions). The counting detector usually consists of a material which emits light when a particle goes through it, and a photon sensor which converts the light into an electrical signal. The most common combination is a plastic scintillator, coupled with a photo-multiplier tube (PMT). Standard PMTs, however, are susceptible to magnetic fields. In a Penning trap environment, there is an unavoidable stray field of up to a few 100 G. For the external annihilation detector for ALPHA, we use triple shields for our PMTs with mu-metal and iron shielding. Such a PMT is coupled to a relatively large scintillator (60 cm x 40 cm x 1 cm), placed outside the main solenoid magnet. Two scintillator-PMT systems are used as a coincidence pair, which reduces noise. The background is mainly due to cosmic rays, typically 20 Hz per coincidence pair. For our first run in 2006 we had 2 such pairs. The coverage was doubled for the 2007 run. We are currently constructing 2 more pairs, making the total to be 6 coincidence pairs, or 12 scintillator-PMT systems. Fig. 2 (left) illustrates an example of early physics results obtained in 2006 with this external annihilation detector.

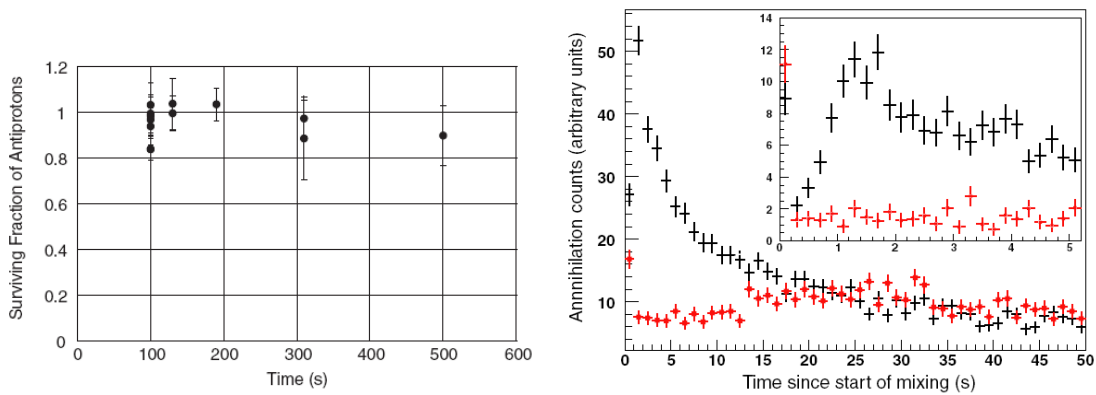

**FIGURE 2.** (Left) Examples of physics results obtained with the external annihilation detector, indicating survival of antiprotons in an octupole field [4]. (Right) Results with the internal annihilation detector, demonstrating antihydrogen production at 1 T field [5].

# Internal Annihilation Detector

With the background level of tens of Hz, the external annihilation detector is not well suited for detecting small signals. The antihydrogen annihilation signal during the mixing of antiprotons and positrons is one such example. In this case, it would be preferable to place the detector inside the bore of the solenoid magnet, close to the Penning trap walls, in order to increase the signal to background. The strong magnetic field and the space restrictions make it difficult to use a PMT as a photon sensor. The Si strip detector we are developing (see below) will work in such an environment, but delays in the delivery by the Si manufacturer forced us to develop an alternative internal annihilation detection system on a very short time scale.

Our initial attempt was based on a design similar to that used for the electromagnetic calorimeter [34] developed for the KOPIO rare kaon decay experiment. The system consisted of a scintillator with extruded holes through which wave-shifting fibers were run to bring the photons out. Unfortunately, the light yield from the KOPIO system was not sufficient for our application. Instead, we developed a system which coupled a scintillator directly to an avalanche photodiode (APD), as shown in Fig. 3. The main volume of the scintillator has a dimension of 30 cm x 7 cm x 2.5 cm. One of the sides of the volume is extended and tapered to match the 8 mm x 8 mm active area of the APD. Both the main volume and the tapered volume are machined out of the same piece of scintillator material. A low noise pre-amplifier has been developed to read out the APD signal. At a typical operating voltage of 1.8 kV, the APD has a gain of about 500. Fig. 2 (right) shows physics data taken in 2006 using an initial system of four scintillator-APD modules.

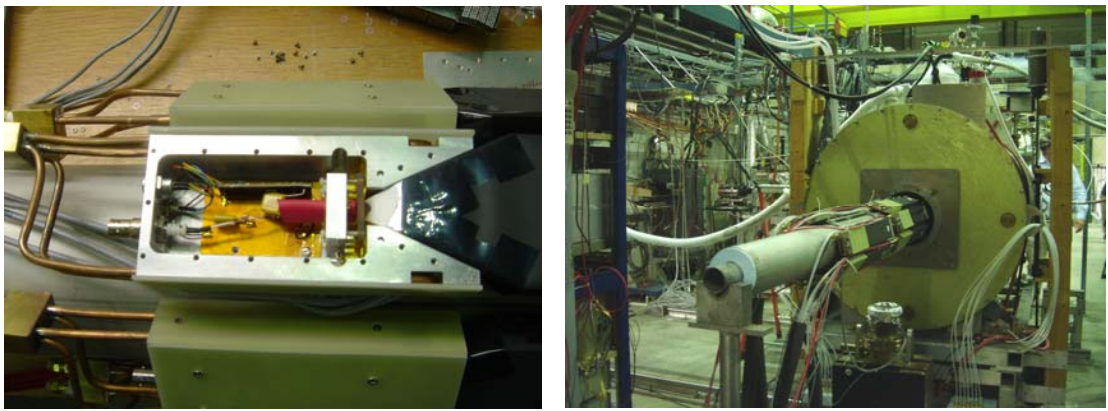

**FIGURE 3**. (Left) Internal annihilation detector assembly. The tapered scintillator is directly coupled to an APD. (Right) Installation of the internal detector into the ALPHA solenoid magnet.

The internal detector system was substantially improved for the 2007 run. The detector assembly was redesigned and rebuilt, which allowed for a better noise insulation as well as improved mechanical strength. The new system allows temperature control of the APD to 0.1 degree C using a water cooling system, which stabilizes the gain of the APD. A total of up to 14 scintillator-APD modules surrounded the Penning traps to provide a high solid angle coverage. Physics results obtained with this new system in the 2007 run were recently published in Ref. [6]. It has also been used to search for trapped antihydrogen as described in [35]

## Antiproton Beam Detectors

ALPHA's beam detection systems use techniques similar to those developed for ATHENA, but adopted specifically for the ALPHA application. The spatial profile of the antiproton beam is measured with a segmented Si detector of a small thickness (60 μm). The detector, operated at a few 10's of Kelvin, is biased typically at -75 V, 10 times higher than the depletion voltage. This is necessary since a pulse of $10^7$ antiprotons generates a large number of electron-hole pairs in Si, some of which recombine unless there is a large electric field to sweep out the charge carriers. We are currently investigating the possibility of a new beam detector based on a CVD diamond film, with even smaller thickness and improved spatial resolution.

The antiproton beam intensity is measured with a scintillator coupled to a hybrid photodiode, a system similar to that discussed in [36].

## Si Vertex Detector

One of the important design features of the ALPHA apparatus is the capability to imagine annihilations by means of a Si vertex detector. As demonstrated in ATHENA [37,38], this is an extremely powerful tool, not only for detecting antihydrogen, but also for diagnosing related plasma processes. Also, for the initial detection of trapped antihydrogen and the subsequent spectroscopy experiments, it will probably necessary to search for very rare events in the presence of backgrounds such as cosmic rays. The high background rejection capability from a position sensitive detector is likely be important for the first detection of trapped antihydrogen and its spectroscopy. This is why we have made a great effort to incorporate annihilation imaging capability compatible with a neutral atom trap. See [39] for conceptual discussions of the ALPHA detector.

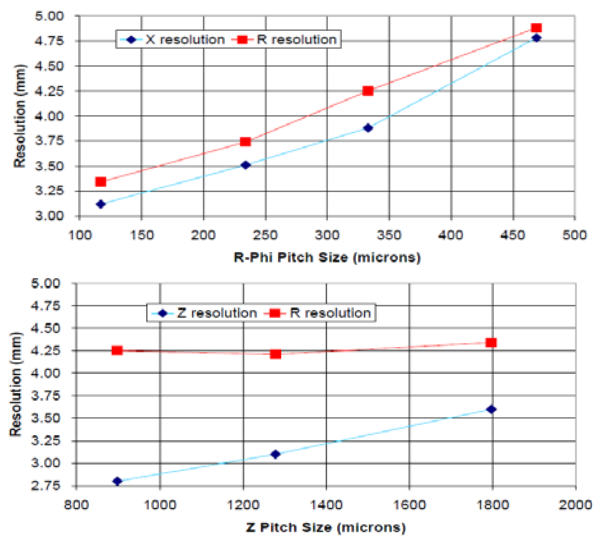

**FIGURE 4**. Monte Carlo simulation studies of the expected resolutions for reconstructed vertices as a function of R-phi and Z pitch sizes. See the text for details [40].

Since the early studies [39], detailed Monte Carlo simulations have been performed in order to determine the design parameters of the ALPHA vertex detector (see, e.g, Fig. 4) [40]. Our Si module has a dimension of 230 mm x 60 mm, and is double-sided. Strip pitches are 900 μm and 230 μm along the z- (along the solenoid field) and phi-directions, respectively. A total of 60 such modules will be used to surround the trapping region in 3 layers. Our simulations show that it is essential to have 3 layers of Si detectors in order to have a useful reconstruction capability, as opposed to the ATHENA detector, which had only 2 layers. The main difference in ALPHA is the increased scattering material thickness and the larger distance from the trap walls, both dictated by the requirements for the neutral (anti-atom) trap.

The data acquisition system to read out and record a total of 30,000 channels of the ALPHA Si detector features a custom-made high density flash ADC (VF48), and also a custom-made trigger and control module (TTC). Si sensors and front-end ASIC chips are connected via Front-end Repeater Cards to the ADCs. The MIDAS software framework controls the readout sequence via a VME controller.

## Si Detector Commissioning

In the final 3 weeks of the 2007 run, we commissioned the first Si modules in the antihydrogen experiment environment at the AD. Two groups of 3 layer Si modules (see Fig. 5), constituting 1/10 of the full detector system, were placed in the trapping region of the ALPHA apparatus. We would have been satisfied with simply detecting a few pion tracks with the detector, but we have managed to reconstruct antiproton annihilation vertices (despite the low efficiency due to the small coverage), and obtain some physics results during our commissioning. Figure 5 illustrates examples of cosmic events, as well as antiproton annihilation events, both detected by the Si detector and reconstructed via a vertex analysis software routine.

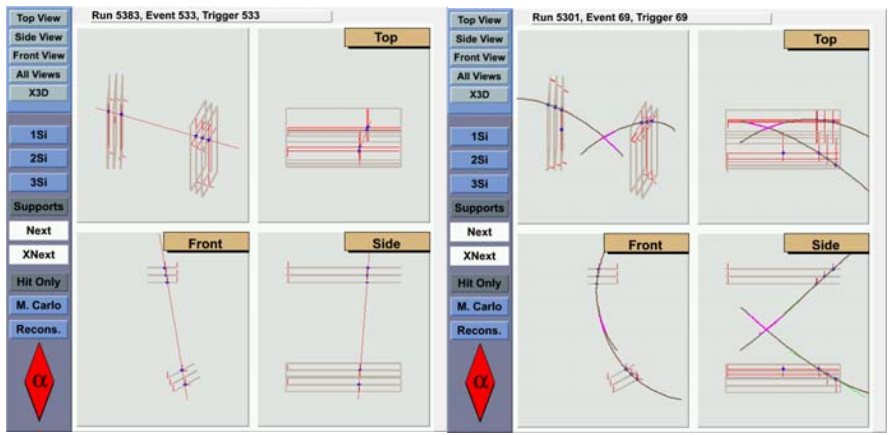

**FIGURE 5**. Reconstructed tracks from a cosmic ray (B = 0 T) and antiproton annihilation (1 T).

The axial (along the solenoid field) distribution of antiproton annihilations is shown in Fig. 6, with the Penning trap potential well moved to different positions. The annihilations follow the trap wells as expected, demonstrating the basic functionality of the entire chain of the Si detector system, starting from the Si sensors, the front-end electronics, the data acquisition system, and the vertex reconstruction routine (including the calibration and the alignment).

Figure 7 shows the axial distribution of antiproton annihilation vertices during mixing of antiprotons and positrons in a uniform 1 T field, a condition similar to that for Fig. 2 (Right). A peak at the position of the positron plasma is characteristic of antihydrogen annihilations (c.f. Ref. [41]), corroborating the production of anti-hydrogen at 1 T, which we reported earlier in Ref. [5].

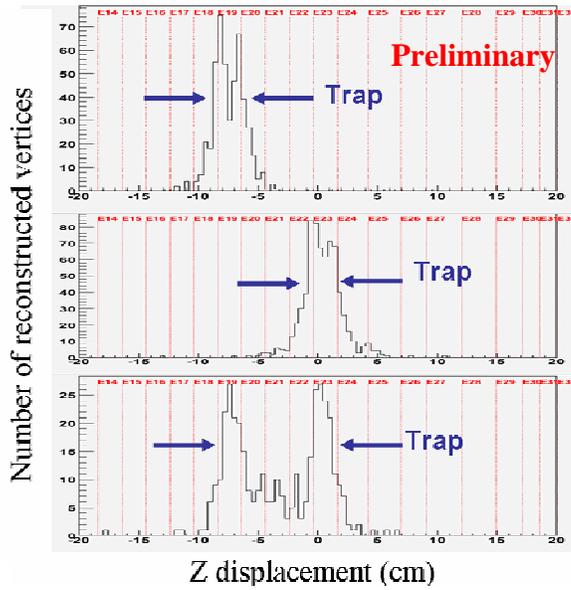

**FIGURE 6**. Axial position of the reconstructed annihilation vertices, when the Penning trap potential walls (indicated with arrows) are moved. Vertical lines indicate the position of the trap electrodes.

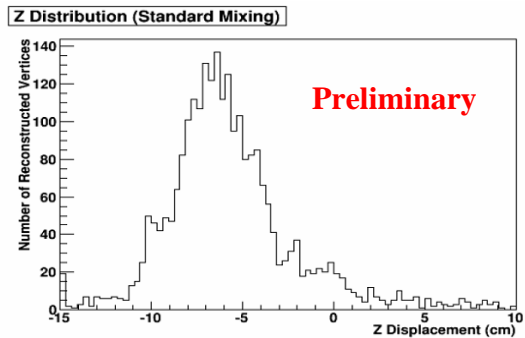

**FIGURE 7**. Axial distribution of the annihilation vertices during antiproton-positron mixing, indicating the production of antihydrogen. The positron plasma is centered near -6 cm.

New physics information, although still preliminary, is obtained with the partial detector during our study of antiproton dynamics in an octupole magnetic field. The axial distribution of antiproton annihilation shown in Fig. 8 (Left) is indicative of the so-called ballistic loss of antiprotons in an octupole field [6]. The larger the axial excursion of the antiproton in an octupole field, the more likely it is that the trajectory of the particle coincides with the trap wall. This leads to the observed concentration of annihilations at both edges of the trapping potential.

Furthermore, the ballistic loss model predicts that antiproton losses are azimuthally peaked four-folds at trap edges, and that these four peaks are shifted by 45 degrees between the edges [6]. The Preliminary z-phi distributions shown in Fig. 8 (Right) provides an indication of such peaks. Note that due to the limited solid angle coverage of the Si modules, the reconstruction efficiency is not uniform. It should be stressed that Figs. 8, while consistent with the ballistic model predictions, are not definitive proof for it, and further studies with a full detector are required. Nonetheless, these examples highlight the power of the annihilation imaging detector, even with 1/10 of the full system, when studying antiproton processes in a new condition. We are currently preparing to install more than 1/2 of the full system for the 2008 run. It is hoped that the full detector system will be completed by the end of 2008.

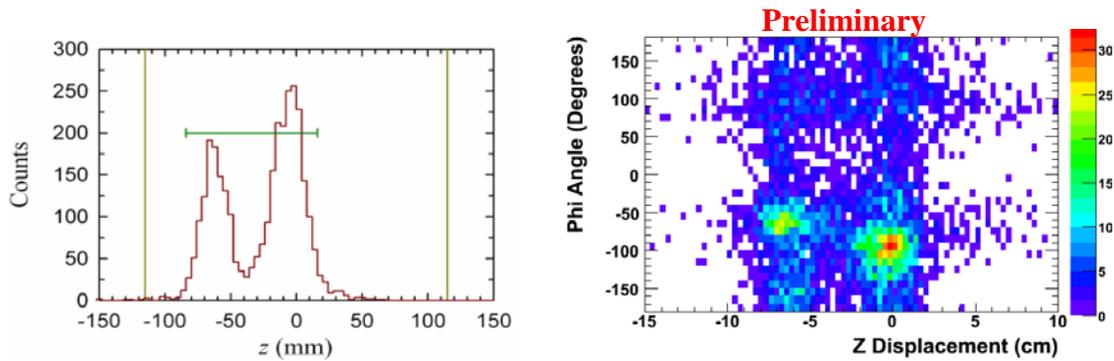

**FIGURE 8**. (Left) Axial distribution of antiproton annihilations when the octupole field is ramped up. Reproduced from Ref. [6]. (Right) The azimuthal, as well as the axial annihilation distributions for the same data. Both plots are consistent with the ballistic model of antiproton loss in an octupole field.

## SUMMARY

In this article we have discussed some aspects of the ALPHA experiment related to particle physics techniques and ideas. We reviewed the fundamental physics case for antihydrogen studies and their potential physics reach, emphasizing the possible sensitivity to Planck scale phenomena once antihydrogen trapping is achieved. We have also discussed some of the particle detection techniques used in ALPHA. Preliminary results for a commissioning run of a partial system of a Si vertex detector were presented, which highlights the power of an annihilation vertex detection capability in antihydrogen studies.

## ACKNOWLEDGMENTS

We gratefully acknowledge the technical support by TRIUMF and the University of Liverpool for detector development, design and construction. In particular, we are indebted to Pierre Amaudruz, Daryl Bishop, Zlatko Bjelic, Travis Howland, Kevin Langton, and Peter Vincent of TRIUMF, Jean-Pierre Martin of Montreal, and David Seddon, Jim Thornhill, and David Wells of the University of Liverpool. We also thank Pablo Genoa, Lawrence Posada, and Alberto Rotondi for their help in the early stages of Si simulations. This work is supported in part by by CNPq, FINEP (Brazil), NSERC, NRC/TRIUMF (Canada), FNU (Denmark), ISF (Israel), MEXT, RIKEN (Japan), EPSRC and Leverhulme Trust (UK) and DOE (USA).